\documentclass{aa}  
\usepackage{natbib}
\usepackage{natbib,twoopt}
\usepackage{graphicx}
\usepackage{txfonts}
\usepackage[para,online,flushleft]{threeparttable}
\bibpunct{(}{)}{;}{a}{}{,}

\newcommand{\kms}{km~s$^{-1}$}
\newcommand{\teff}{$T_{\rm eff}$}
\newcommand{\logg}{log~$g$}
\newcommand{\mh}{[M/H]}
\newcommand{\am}{[$\alpha$/M]}
\newcommand{\cm}{[C/M]}
\newcommand{\vmicro}{$v_{\rm micro}$}
\newcommand{\mum}{$\mu$m}
\newcommand{\ebv}{E(B$-$V)}

\begin{document}

\title{The updated BOSZ synthetic stellar spectral library}

\author{
Szabolcs~M{\'e}sz{\'a}ros\inst{1,2}, 
Ralph~Bohlin\inst{3}, 
Carlos Allende Prieto\inst{4,5}, 
Borb{\'a}la Cseh\inst{2,6,7}, 
J{\'o}zsef Kov{\'a}cs\inst{1,2,8}, 
Scott~W.~Fleming\inst{3}, 
Zolt{\'a}n Dencs\inst{1,2}, 
Susana~Deustua\inst{3,9}, 
Karl~D.~Gordon\inst{3}, 
Ivan Hubeny\inst{10}, 
Gy{\"o}rgy Mez{\H{o}}\inst{6,7}, 
M{\'a}rton Truszek\inst{11}
}

\institute{
ELTE E\"otv\"os Lor\'and University, Gothard Astrophysical Observatory, 9700 Szombathely, Szent Imre H. st. 112, Hungary
\and
MTA-ELTE Lend{\"u}let "Momentum" Milky Way Research Group, Szent Imre h. u. 112., Szombathely H-9700, Hungary 
\and
Space Telescope Science Institute, 3700 San Martin Drive, Baltimore, MD 21218, USA
\and
Instituto de Astrofísica de Canarias (IAC), E-38200 La Laguna, Tenerife, Spain
\and
Departamento de Astrofísica, Universidad de La Laguna (ULL), E-38206 la Laguna, Tenerife, Spain
\and
Konkoly Observatory, HUN-REN Research Centre for Astronomy and Earth Sciences (CSFK), Konkoly Thege M. út 15-17, 1121, Budapest, Hungary
\and
CSFK, MTA Centre of Excellence, Budapest, Konkoly Thege Mikl\'os \'ut 15-17., H-1121, Hungary
\and 
HUN-REN–ELTE Exoplanet Systems Research Group, Szent Imre h. u. 112., Szombathely H-9700, Hungary 
\and
Sensor Science Division, National Institute of Standards and Technology, Gaithersburg, MD 20899-8441, USA
\and 
The University of Arizona, Steward Observatory, 933 North Cherry Avenue, Tucson, AZ 85719, USA
\and
Department of Experimental Physics, Institute of Physics, University of Szeged, Dóm tér 9, 6720, Szeged, Hungary
}


\abstract 
{The modeling of stellar spectra of flux standards observed by the \textit{Hubble} and \textit{James Webb} space telescopes requires a large synthetic spectral library that covers a wide atmospheric parameter range.}
{The aim of this paper is to present and describe the calculation methods behind the updated version of the BOSZ synthetic spectral database, which was originally designed to fit the CALSPEC flux standards. These new local thermodynamic equilibrium (LTE) models incorporate both MARCS and ATLAS9 model atmospheres, updated continuous opacities, and 23 new molecular line lists.}
{The new grid was calculated with Synspec using the LTE approximation and covers metallicities \mh \ from $-$2.5 to 0.75~dex, \am \ from $-$0.25 to 0.5~dex, and \cm \ from $-$0.75 to 0.5~dex, providing spectra for 336 unique compositions. Calculations for stars between 2800 and 8000~K use MARCS model atmospheres, and ATLAS9 is used between 7500 and 16,000~K.}
{The new BOSZ grid includes 628,620 synthetic spectra from 50 nm to 32 \mum \ with models for 495 \teff\ -- \logg\ parameter pairs per composition and per microturbulent velocity. Each spectrum has eight different resolutions spanning a range from R = 500 to 50,000 as well as the original resolution of the synthesis. The microturbulent velocities are 0, 1, 2, and 4 \kms.}
{The new BOSZ grid extends the temperature range to cooler temperatures compared to the original grid because the updated molecular line lists make modeling possible for cooler stars. A publicly available and consistently calculated database of model spectra is important for many astrophysical analyses, for example spectroscopic surveys and the determination of stellar elemental compositions.}


\titlerunning{New BOSZ grid}
\authorrunning{M{\'e}sz{\'a}ros et al. 2024}
\maketitle

\section{Introduction}

The design of the \textit{James Webb }Space Telescope (JWST) absolute flux calibration program follows the absolute calibration methods used for the \textit{Hubble} Space Telescope (HST) and directly ties measured fluxes to laboratory standards \citep{2022AJ....163..267G}. Nonlocal thermodynamic equilibrium (NLTE) models of three primary hot white dwarfs define the flux calibrations for HST. These theoretical spectral energy distributions (SEDs) provide the relative flux versus wavelength, and only the absolute flux level must be reconciled between the measured absolute flux of Vega in the visible and the Midcourse Space Experiment values for Sirius in the mid-IR \citep{2014PASP..126..711B, 2020AJ....160...21B}. Fitting of NLTE models to Balmer line profiles determines the \teff\ and \logg\ of the three primary white dwarfs. To produce secondary flux standards for calibrating JWST, model spectra with IR coverage to 32~\mum\ are then fit to HST observations shortward of 2.5~\mum.

While many high-resolution  theoretical spectral libraries \citep{2008A&A...486..951G, 2004A&A...417.1055Z, 2014MNRAS.440.1027C} based on either ATLAS9 \citep{1979ApJS...40....1K} or MARCS \citep{2008A&A...486..951G} model atmospheres already existed, the flux calibration of JWST required a grid that extended to 30 $\mu$m. Our previous work, the BOSZ (named using letters from the names of the first two authors) synthetic spectral grid \citep{2017AJ....153..234B}, was created to fit the secondary flux standards from the CALSPEC database\footnote{http://www.stsci.edu/hst/instrumentation/reference-data-for-calibration-and-tools/astronomical-catalogs/calspec}.

This grid utilized the APOGEE-ATLAS model atmosphere database \citep{2012AJ....144..120M} of ATLAS9 models \citep{1979ApJS...40....1K} that was originally calculated for the SDSS-III APOGEE survey \citep{2017AJ....154...94M}. The high-resolution spectra were synthesized with SYNTHE \citep{1981SAOSR.391.....K} using the Linux-ported version \citep{2004MSAIS...5...93S} and the \citet{2005ASPC..336...25A} solar reference abundances. With a total of 336 different compositions, as well as 14 \mh, 4 \am, and 6 \cm \ values that cover the majority of observed stellar atmospheric parameters, this grid was the largest such theoretical spectrum database available online. The spectra spanned the wavelength range of 100 nm – 32 $\mu$m with vacuum wavelengths using an atomic line list compiled by Robert Kurucz in 2016, and with line lists of 12 molecules from Kurucz's website\footnote{http://kurucz.harvard.edu/}. Each composition had 415 models for a total of 139,440 theoretical spectra in the final grid. 

\citet{2018A&A...618A..25A} used more recent line and continuum opacities than those available for Kurucz’s
SYNTHE code and synthesized a large spectral grid with ASS$\epsilon$T \citep{2008ApJ...680..764K, 2009AIPC.1171...73K} between 120 and 6500 nm using the APOGEE-ATLAS model atmospheres \citep{2012AJ....144..120M}. While their spectral library does not cover the entire JWST wavelength range, important updates were made in the treatment of radiative transfer and continuous opacities. Both \citet{2017AJ....153..234B} and \citet{2018A&A...618A..25A} used older molecular line lists compiled by Robert Kurucz; but recently, the ExoMol project \citep{2017JQSRT.203..490B, 2017JQSRT.187..453B} published new molecular line lists that enabled further refinements to the online spectral databases.

The goal of this paper is to present the updated BOSZ grid, in which the temperature range is extended down to 2800~K, continuum opacities are updated from \citet{2018A&A...618A..25A}, and an extended molecular line list for the atmospheres of cool stars is included. The paper is structured as follows: Section 2 details the model atmosphere database behind the new grid, in Section 3 the atmospheric parameters of the new spectral library and the synthesis are described, and in Section 4 the old and new BOSZ grids are compared for three CALSPEC stars.

\section{Model atmosphere assumptions}

\subsection{MARCS and ATLAS9}

One of the most significant updates in the new calculations is the replacement of ATLAS9 \citep{1979ApJS...40....1K} model atmospheres with MARCS \citep{2008A&A...486..951G} for stars below 8000~K, while above 8000~K ATLAS9 models from \citet{2012AJ....144..120M} using \citet{2005ASPC..336...25A} as solar reference remain the same\footnote{https://www.iac.es/proyecto/ATLAS-APOGEE/}. The cool end of the BOSZ grid now extends down to 2800~K. The MARCS grid used here was originally computed for the derivation of atmospheric parameters and abundances released by APOGEE in the 16th data release of SDSS \citep{2020ApJS..249....3A}. The description of these models is explained by \citet{2020AJ....160..120J}, and the grid is on the SDSS website\footnote{https://data.sdss.org/sas/dr17/apogee/spectro/speclib/atmos/marcs/}. 
The solar reference abundances are from \citet{2007SSRv..130..105G}, and the $\alpha$ elements are defined as O, Ne, Mg, Si, S, Ar, Ca, and Ti. An important upgrade is the inclusion of C$_{\rm 2}$H$_{\rm 2}$ and C$_{\rm 3}$ line opacities in the MARCS models, which make their atmospheric structure more realistic. A detailed analysis of the effect of these molecules on the atmospheres is presented by \citet{2008A&A...486..951G}.

\begin{figure*}                
\centering
\includegraphics[width=6.9in,angle=0]{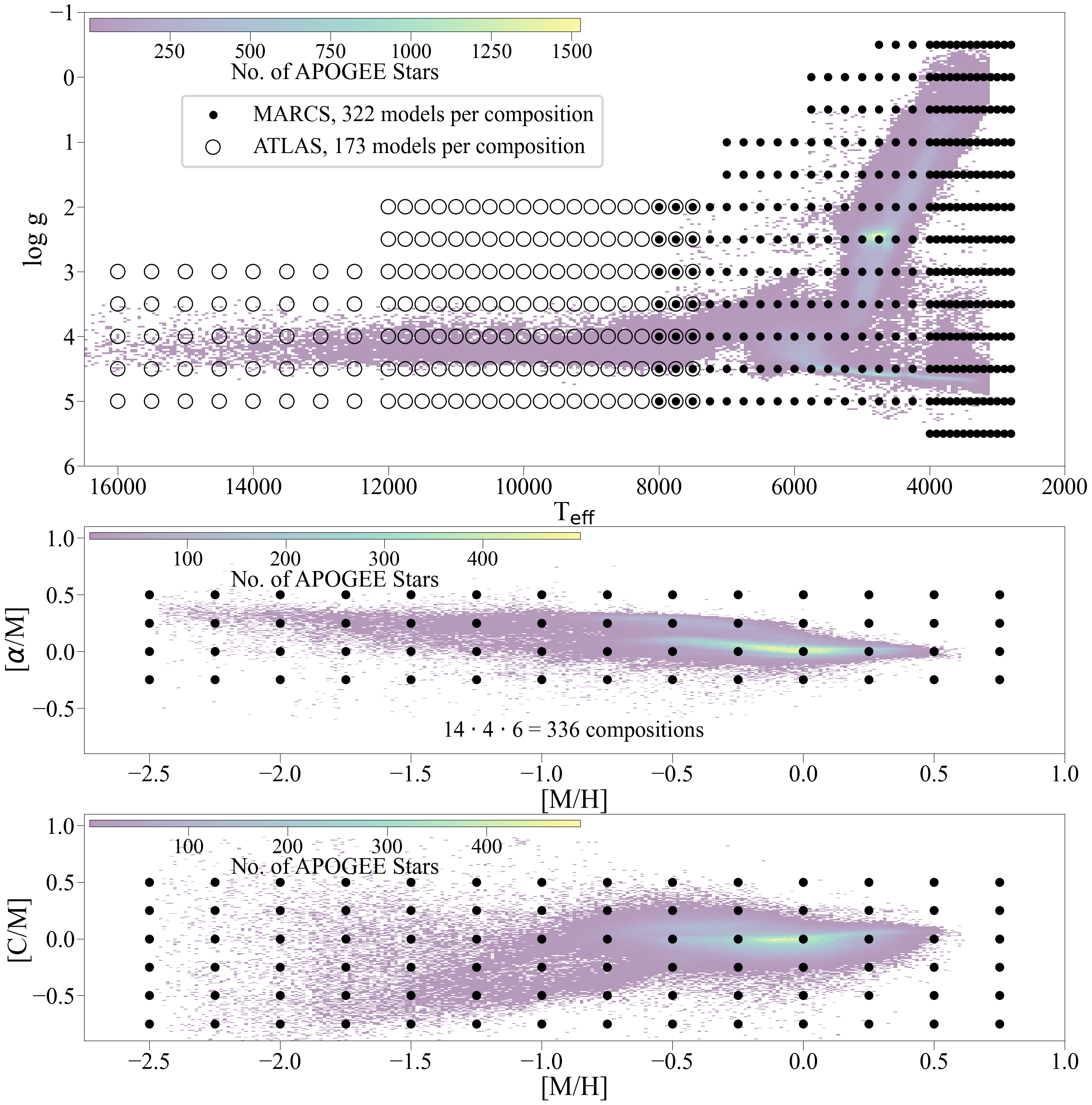}
\caption{The parameter space covered by the new BOSZ grid. Top panel: \teff \ $-$ \logg \ diagram of the new BOSZ grid. Middle
panel: \am \ as a function of \mh. Bottom panel: \cm \ as a function of \mh.
There are 336 different chemical compositions included in the database, and each
composition has 322 MARCS and 173 ATLAS9 model atmospheres. The star density from APOGEE
DR17 \citep{2022ApJS..259...35A} is the colored background and shows that the
selected parameters cover most of the stars that are observed in the Milky Way.}
\label{grid}
\end{figure*}

This new MARCS grid is substantially larger than the MARCS models described in \citet{2012AJ....144..120M}. This paper includes only the subset of the chemical element perturbations that are relevant for most of the stars observed by the large high-resolution  spectroscopic sky surveys. The chemical composition of the new BOSZ grid is listed in Table~1 and shown in the bottom two panels of Figure~\ref{grid}. Metallicity is varied between $-$2.5 and 0.75~dex, the \am \ between $-$0.25 and 0.5~dex, and \cm \ between $-$0.75 and 0.5~dex with 0.25~dex steps for all three parameters. These chemical variations are the same for all \teff, \logg \ pairs in the entire grid.

Within each composition, seven different mini-grids are defined based on the \teff \ and \logg \ listed in Table~2. Below 4000~K, the step size is 100~K, between 4250 and 12000~K steps are 250~K, and above 12000~K 500~K. Unlike the old BOSZ database, models with temperature higher than 16000~K are not included due to the lack of the necessary NLTE effects in ATLAS9. For this reason, \citet{2022AJ....164...10B} recently replaced the BOSZ models with NLTE ones. The step size of \logg \ is 0.5~dex everywhere, but the minimum value of surface gravity increases with increasing temperature to roughly follow the Eddington-limit. In the MARCS model atmospheres, spherical geometry is used between \logg \ = $-$1 and 3 with \vmicro \ = 2 \kms; and plane-parallel geometry is assumed between \logg= 3.5 and 5.5~dex with \vmicro \ = 1 \kms. All ATLAS9 models are plane-parallel model atmospheres with \vmicro \ = 2 \kms. The \teff$-$\logg \ diagram of our selected parameters appears in the top panel of Figure~\ref{grid}. Spectra are synthesized from both ATLAS9 and MARCS model atmospheres between 7500 and 8000~K.

\begin{scriptsize}
\begin{table}
\caption{Chemical composition of all MARCS and ATLAS9 spectra.}
\begin{tabular}{p{1.82cm}p{1.82cm}p{1.82cm}p{1.82cm}}
\hline
Par. & Min. & Max. & Step \\
\hline
\mh & $-$2.5 & 0.75 & 0.25 \\
\am & $-$0.25 & 0.5 & 0.25 \\
\cm & $-$0.75 & 0.5 & 0.25 \\
\hline
\end{tabular}
\begin{tablenotes}
\small
\end{tablenotes}
\end{table}
\end{scriptsize}

\begin{scriptsize}
\begin{table}
\caption{Temperature and surface gravity range of the BOSZ grid.}
\begin{tabular}{p{0.65cm}p{0.65cm}p{0.65cm}p{0.65cm}p{0.65cm}p{0.65cm}rp{0.65cm}}
\hline
T$_{\rm eff}$ Min. & T$_{\rm eff}$ Max. & T$_{\rm eff}$ Step & $\log$~g Min. & $\log$~g Max. & $\log$~g Step & n\\
\hline
MARCS &  &  &  &   &  & \\
\hline
2800 & 4000 & 100 &  -0.5 & 5.5 & 0.5 & 203\,928\\
4250 & 4750 & 250 & -0.5 & 5 & 0.5 & 43\,456 \\
5000 & 5750 & 250 & 0 & 5 & 0.5 & 54\,484\\
6000 & 7000 & 250 & 1 & 5 & 0.5 & 57\,712\\
7250 & 8000 & 250 & 2 & 5 & 0.5 & 36\,528\\
\hline
ATLAS9 &  &  &  &  &  &  & \\
\hline
7500 & 12000 & 250 & 2 & 5 & 0.5 & 178\,752\\
 12500 & 16000 & 500 & 3 & 5 & 0.5 & 53\,760
\end{tabular}
\begin{tablenotes}
\small
\item All models are calculated with four microturbulent velocity values: \vmicro = 0, 1, 2, 4 \kms; and 8 resolutions: original, 500, 1000, 2000, 5000, 10\,000, 20\,000, and 50\,000.
\end{tablenotes}
\end{table}
\end{scriptsize}

\begin{scriptsize}
\begin{table}
\caption{Parameters of the missing MARCS models.}
\begin{tabular}{p{1.37cm}p{1.37cm}p{1.37cm}p{1.37cm}p{1.37cm}}
\hline
\teff & \logg & \mh & \am & \cm \\
\hline
2800 & +0.0 & +0.00 & +0.00 & +0.50 \\
2800 & +0.0 & +0.00 & +0.25 & -0.25 \\
2800 & +0.0 & +0.00 & -0.25 & +0.25 \\ 
2800 & +0.0 & +0.00 & +0.25 & -0.75 \\
2800 & +0.0 & +0.00 & +0.50 & -0.25 \\
\end{tabular}
\begin{tablenotes}
\small
\item The full version of the table is available at the CDS.
\end{tablenotes}
\end{table}
\end{scriptsize}

\subsection{Missing models}

While both model atmosphere calculation methods can suffer from convergence problems, the ATLAS9 grid is complete between 8000 and 16000~K. Unfortunately, the MARCS model atmospheres do not always fully converge at low temperatures and low or very high surface gravities, or for unconventional chemical compositions. This issue is most apparent at the edges of the parameter space. Out of the 108\,192 MARCS files, 9165 did not converge (8.5\%), and their parameters are listed in Table~3. Most of the un-converged models (i.e., holes) are near the top of the red giant branch as only 24\% of the \logg \ = $-$0.5 cases could be computed.

\begin{scriptsize}
\begin{table*}
\caption{Molecular line lists.}
\begin{tabular}{p{1.3cm}p{8.1cm}p{7.6cm}}
\hline
Molecule & File Path & Reference \\
\hline
AlH & molecules/alh/alhax.asc, 
molecules/alh/alhxx.asc & \citet{2018MNRAS.479.1401Y} \\
AlO & molecules/alo/alopatrascu.asc & \citet{2015MNRAS.449.3613P} \\
C$_2$ & molecules/c2/c2dabrookek.asc, 
linelists/linesmol/c2ax.asc & \citet{2013JQSRT.124...11B} \\
& linelists/linesmol/c2ba.asc, 
linelists/linesmol/c2ea.asc & \\
CaH & molecules/cah/cah.asc & \citet{2003JChPh.118.9997W, 2012MNRAS.425...34Y, 2012JQSRT.113...67L,  2013JMoSp.288...46S} \\
CaO & molecules/cao/caoyurchenko.asc & \citet{2016MNRAS.456.4524Y} \\
CH & molecules/ch/chmasseron.asc & \citet{masseron1} \\
CN & molecules/cn/cnaxbrookek.asc, 
molecules/cn/cnbxbrookek.asc & \citet{2014ApJS..210...23B} \\
& molecules/cn/cnxx12brooke.asc & \\
CO & molecules/co/coax.asc, 
linelists/linesmol/coxx.asc & http://kurucz.harvard.edu/ \\
CrH & molecules/crh/crhaxbernath.asc & \citet{2001JChPh.115.1312B, 2002ApJ...577..986B} \\
FeH & molecules/feh/fehfx.asc & \citet{2003ApJ...594..651D} \\
H$_2$ & linelists/linesmol/h2.asc, 
linelists/linesmol/h2bx.asc & http://kurucz.harvard.edu/ \\
& linelists/linesmol/h2cx.asc, 
linelists/linesmol/h2xx.asc & \\
H$_2$O & Synspec addition & \citet{2017JQSRT.203..490B, 2017JQSRT.187..453B, 2021arXiv210402829H} \\
MgH & molecules/mgh/mgh.asc & \citet{2007JPCA..11112495S, 2012MNRAS.425...34Y, 2013MNRAS.432.2043G, 2013ApJS..207...26H} \\
MgO & molecules/mgo/mgodaily.asc & \citet{2002JMoSp.214..111D} \\
NaH & molecules/nah/nahrivlin.asc & \citet{2015MNRAS.451..634R} \\
NH & linelists/linesmol/nh.asc & http://kurucz.harvard.edu/ \\
OH & molecules/oh/ohupdate.asc & http://kurucz.harvard.edu/ \\
OH+ & molecules/ohplus/ohplusax.asc, 
molecules/ohplus/ohplusxx.asc & \citet{2018ApJ...855...21H} \\
SiH & molecules/sih/sihaxsightly.asc, 
molecules/sih/sihxxsightly.asc & \citet{2018MNRAS.473.5324Y} \\
SiO & linelists/linesmol/sioax.asc, 
linelists/linesmol/sioex.asc & http://kurucz.harvard.edu/ \\
& linelists/linesmol/sioxx.asc &  \\
TiH & molecules/tih/tih.asc & \citet{2005ApJ...624..988B} \\
TiO & Synspec addition & \citet{2019MNRAS.488.2836M, 2021arXiv210402829H} \\
VO & molecules/vo/vomyt.asc & \citet{2016MNRAS.463..771M} \\
\end{tabular}
\begin{tablenotes}
\small
\item The file path comes after http://kurucz.harvard.edu/. H$_2$O and TiO line lists were formatted for Synspec by \citet{2021arXiv210402829H}.
\end{tablenotes}
\end{table*}
\end{scriptsize}

To make a complete grid, these holes were filled by copying other model atmospheres similar in chemical composition to the missing models by \citet{2020AJ....160..120J}. While this would make it possible to have a complete BOSZ grid, we chose not to calculate spectra from these model atmospheres because they have a different chemical composition than what is expected at that position in the grid. A possibility to bypass this issue is to perform interpolation in the model atmospheres as described by \citet{2023A&A...675A.191W}, but it was found by \citet{2013MNRAS.430.3285M} to be less accurate than interpolation of fluxes. However, the iNNterpol code was not yet available when our calculations were made. Spectra related to model atmospheres with parameters from Table~3 are not published. In total, 157\,155 model atmospheres formed the basis of the spectral synthesis, 99\,027 of MARCS, and 58\,128 of ATLAS9.

\section{Spectral synthesis}

\subsection{Radiative transfer and continuum opacities}

The new synthetic BOSZ library is computed with Synspec \citep{2021arXiv210402829H} and its Python wrapper called Synple\footnote{https://github.com/callendeprieto/synple}. After reading the model atmosphere, the code prepares the solution of the equation of state, reads the atomic and molecular line lists to compute the total line opacity, adds the continuum opacity, and then solves the radiative transfer equation. While Synple can handle NLTE models, all spectra described in this work are synthesized under the assumption of local thermodynamic equilibrium (LTE). Synple reads the composition from the model atmosphere file, ensuring consistency between these two calculation steps, that is, ensuring that the solar composition is the same as that of the model atmospheres. 

Partition functions of 29 molecules from the ExoMol project \citep{2012MNRAS.425...21T, 2016JMoSp.327...73T} are implemented in Synspec as described in detail by \citet{2021arXiv210402829H}. All calculations use generic opacities from \citet{2018A&A...618A..25A}, except for molecules (see the next section). There is only a short overview her; all the details of the opacity updates are presented in \citet{2018A&A...618A..25A}. Bound-free opacities of H$^-$, H I, H$_2^+$, CH, and OH and the first two ionization stages of He, C, N, O, Na, Mg, Al, Si, Ca, and Fe are included in the calculations with free-free opacities from H$^-$, H$_2^+$, and He$^-$. The atomic line list is the 2017 version from the Kurucz website\footnote{http://kurucz.harvard.edu/} with the updates to the van der Waals damping constants added by \citet{2018A&A...618A..25A}. The Stark broadening of hydrogen lines were taken from \citet{2009ApJ...696.1755T}.

\subsection{Opacities of molecules}

\begin{figure*}                     
\centering
\includegraphics[width=7.2in,angle=0]{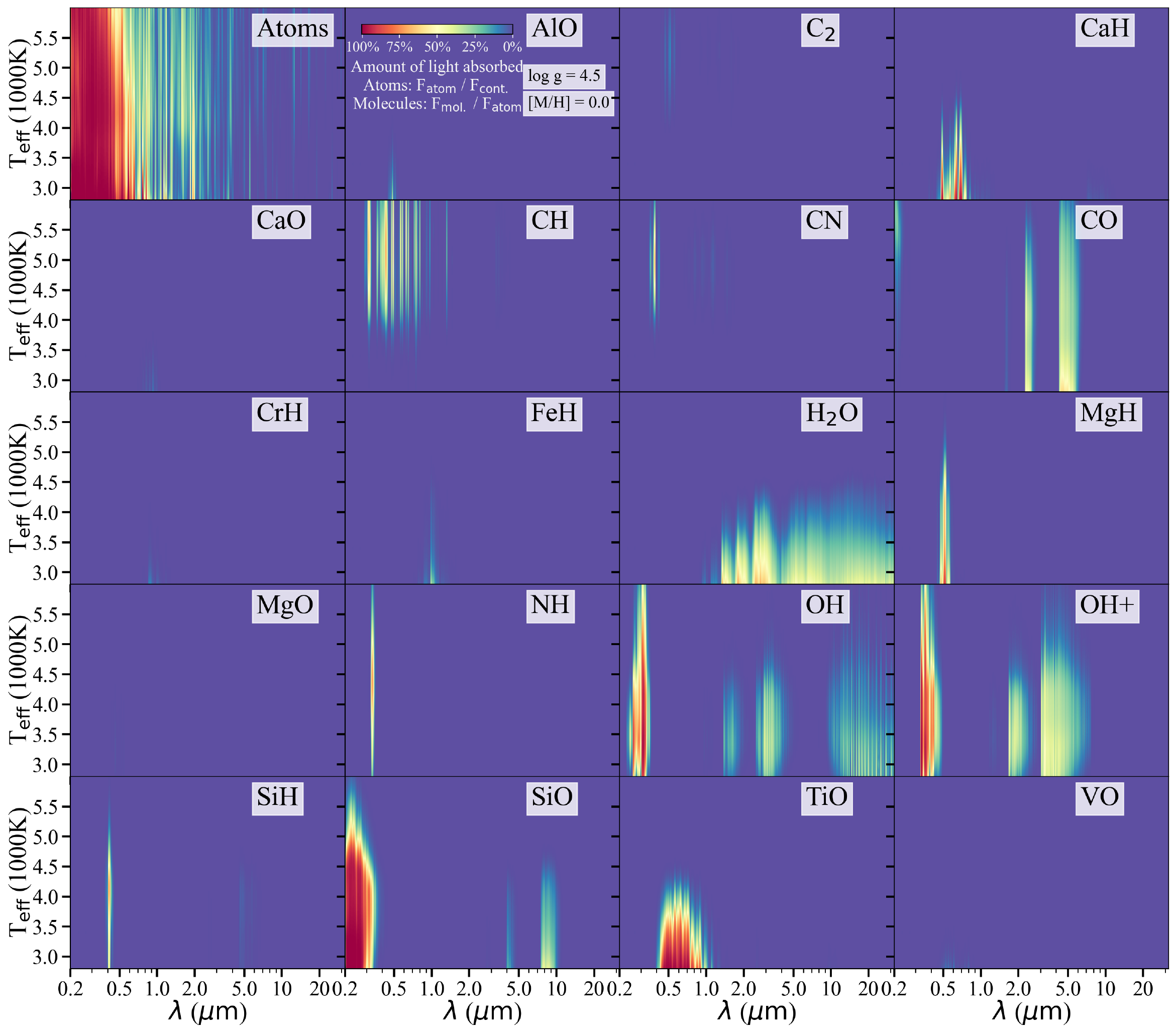}
\caption{Atomic and molecular absorption in dwarf stars with \mh=\am=\cm= 0, \logg \ = 4.5 as a function of wavelength, and \teff. The color indicates the strength of the absorption. The top-left panel shows the atomic absorption only. }
\label{molecules}
\end{figure*}

Significant updates were made to the molecular line list compared to \citet{2017AJ....153..234B} and \citet{2018A&A...618A..25A} by including line lists from the ExoMol project. The improved atom physical data of molecular transitions and increased number of lines makes it possible to compute more accurate synthetic spectra, than with older line lists.

The old BOSZ grid included the following 12 molecules (from the 2015 version of the Kurucz database): C$_2$, CH, CN, CO, H$_2$, H$_2$O, MgH, NH, OH, SiH, SiO, and TiO. This list is expanded here, and altogether, 11 new molecules are considered: AlH, AlO, CaH, CaO, CrH, FeH, MgO, NaH, OH+, TiH, and VO. All but two of the molecules included in the new calculations were compiled by Robert Kurucz using various literature sources, as listed in Table~4. The absorption profiles of 19 molecules appear in Figure~\ref{molecules} as a function of wavelength and \teff. AlH, H$_2$, NaH, and TiH are not shown due to their very weak absorption within our parameter range.

The following molecules are the same in both the old and new calculations: C$_2$, H$_2$, CO, and SiO. However, many previously used molecular line lists have been updated using new line lists calculated as part of the ExoMol project \citep[for references, see Table~4]{2012MNRAS.425...21T, 2016JMoSp.327...73T}. The most important decision was to replace the line lists of H$_2$O and TiO. The H$_2$O line data of the old BOSZ grid were from \citet{1997JChPh.106.4618P}, which is replaced by the one calculated for ExoMol by \citet{2017JQSRT.203..490B, 2017JQSRT.187..453B} and formatted for Synspec by \citet{2021arXiv210402829H}. TiO is also taken from ExoMol database calculated by \citet{2019MNRAS.488.2836M}, instead of \citet{1998FaDi..109..321S}, which was used for the previous BOSZ grid. The original ExoMol TiO line data contained nearly 10$^8$ lines, which was truncated down to the strongest $8.3 \cdot 10^6$ lines, as described in \citet{2021arXiv210402829H}. The partition functions for molecules have also been updated following the ExoMol calculations. 

\begin{figure}                          
\centering
\includegraphics[width=3.54in,angle=0]{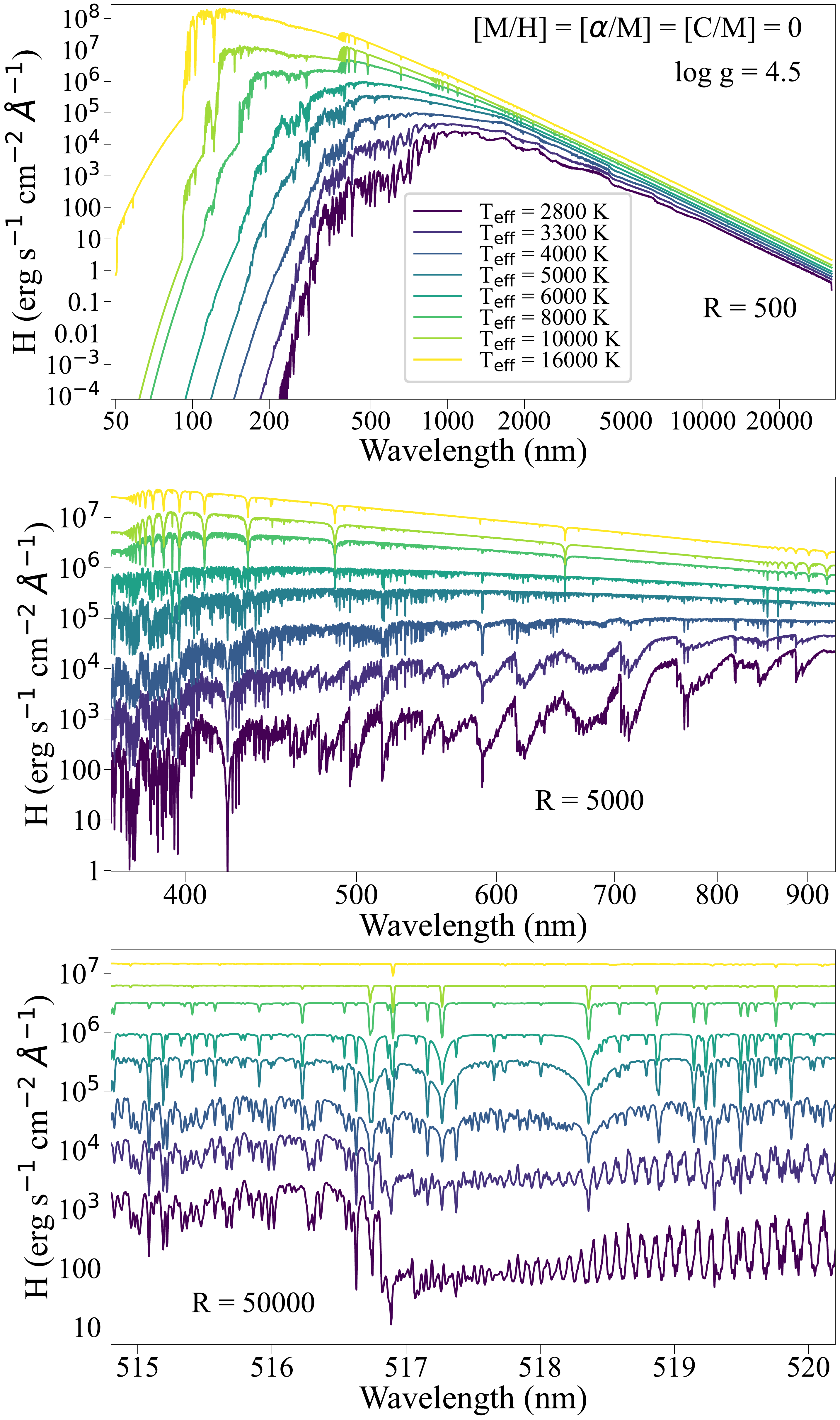}
\caption{Examples of the new BOSZ synthetic spectra as a function of wavelength. Top panel: Example of low-resolution ($R$ = 500) solar composition spectra on the main sequence (\logg \ = 4.5) between 2800 and 16\,000~K. Middle panel: Medium resolution ($R$ = 5000) sequence in the optical and near-IR. Bottom panel: High-resolution  spectra ($R$ = 50\,000) near the Mg triplet lines between 515 and 520 nm.}
\label{bosz}
\end{figure}

Many of the new molecules have significant absorption impacting the atmosphere of stars cooler than 5000~K, as depicted in Figure~\ref{molecules}. These molecules are primarily: CaH, FeH, OH+, and VO. CaH is important in the optical range between 400 and 900~nm below 5000~K, while FeH lines appear in the near IR between 800 and 1700~nm at similar temperatures to CaH. VO lines in the range of 500 and 900~nm are generally weak in the spectra of main sequence stars, but become stronger in the atmosphere of giant stars between below 3500~K. OH+ has wide absorption features between 1.5 and 8 \mum\ that were not accounted in \citet{2017AJ....153..234B}. Another improvement is the inclusion of the extensive list of CH from \citet{masseron1} that affects the spectra of stars between 3500 and 7000~K. 

\subsection{Parameters of the spectral synthesis}

All models were calculated with four microturbulent velocity values, \vmicro = 0, 1, 2, and 4 \kms, which is a significant update from the old BOSZ grid that has only 2 \kms. All spectra span a range from 50 nm to 32 \mum \ using air wavelengths above 200 nm and vacuum wavelength below 200 nm. Due to the immerse calculation time required, the wavelength range was divided into 50 regions and calculated in parallel using the mpsyn routine of Synple. The top panel of Figure~\ref{bosz} illustrates the full wavelength and \teff \ range of solar composition main sequence stars (\logg \ = 4.5). 
Four \vmicro\ calculations for each model atmosphere produce a total of 628\,620 synthetic spectra. The number of spectra available for each mini-grid can be seen in Table~2. The updated BOSZ grid is available via the Mikulski Archive for Space Telescopes (MAST) at the STScI as a High Level Science Product\footnote{https://archive.stsci.edu/prepds/bosz} via https://dx.doi.org/10.17909/T95G68.

The wavelength step of the original synthesis ($\Delta\lambda_o$), automatically chosen in Synple, is calculated the following way:

\begin{equation}
    \Delta\lambda_o = \frac{\lambda_{\rm range}}{c \cdot 0.785} \cdot \sqrt{0.1289^2 \cdot \frac{T_{\rm min}}{100} + \frac{v_{\rm micro}^2}{2}}
,\end{equation}
where $\lambda_{\rm range}$ is the wavelength range, $c$ is the speed of light, and $T_{\rm min}$ is the minimum temperature of the model atmosphere. This step is one third of the thermal and microturbulent broadening at the lowest temperature of the atmosphere for an atomic mass of 100. Thus, each original spectrum has a slightly different sampling and resolution spanning a range roughly between 200\,000 and 600\,000 depending on the main atmospheric parameters. The file name of these original spectra contains the ``rorig'' string and has three columns: the wavelength in $\AA$, the Eddington first moment ($H$, surface brightness) of the spectra, and the continuum. The flux ($F$) at the stellar surface is calculated as

\begin{equation}
F = 4 \pi H
,\end{equation}
where $H$ is in erg~s$^{-1}$ cm$^{-2}$ $\AA^{-1}$.

Synspec solves the radiative transfer equation assuming plane-parallel geometry, irrespective of the geometry of the input model atmosphere, which in the case of MARCS for low gravity stars are spherical models. While the original spectra are included in the database, the following lower resolutions are also provided: $R$ = 500, 1000, 2000, 5000, 10\,000, 20\,000, and 50\,000. The middle panel of Figure~\ref{bosz} shows examples of $R$ = 5000 spectra in the optical and near-IR, while the bottom panel illustrates high-resolution, $R$ = 50\,000 spectra of the Mg Ib triplet lines in the optical.

Each of the seven sets of resolutions has its own equally sampled logarithmic wavelength space for simplicity. To save disk space, the sets of files for the seven lowest resolution values have only the flux and continuum columns with their common wavelengths specified separately. The wavelengths have at least 2 points per resolution element at the lowest temperature (2800~K) for the solar composition. There are 7500 wavelength points at $R$~=~500, which is set as the reference value. Using this approach, the higher temperature spectra are usually over-sampled by more than a factor of 2. The logarithmic wavelength step ($\Delta\lambda_R$) as a function of resolution is calculated with the following equation:

\begin{equation}
\Delta\lambda_R = \frac{\ln(\lambda_{\rm max}) - \ln(\lambda_{\rm min})}{7500 \cdot R / 500}
,\end{equation}
where $\lambda_{\rm max} = 320\,000~\AA$, $\lambda_{\rm min} = 500~\AA$, and $R$ is the resolving power. The wavelength sampling can be downloaded in separate files for each resolution from the BOSZ website at MAST, or each wavelength value for a particular resolution can be calculated with the equation

\begin{equation}
\lambda_{R, i} = e^{ln(500) + \Delta\lambda_R \cdot i}
,\end{equation}
where the $i$ integer starts at 0 and ends with $7500 \cdot R / 500 - 1$.

In FGKM-type stars, the flux is near zero well above 50 nm. To avoid having unreliably small fluxes in the seven sets of lower-resolution spectra, all line flux values are replaced by 0 and continuum fluxes by 1 when the original flux is below 10$^{-6}$~erg~s$^{-1}$ cm$^{-2}$$\AA^{-1}$, which ensures that the same number of rows are in all files of the same resolution. The original output of the synthesis, the highest-resolution spectra (``rorig'' files) retain all fluxes, regardless of their values. 

Line identifications are saved in the ``lineid'' files for \vmicro \ = 0 \kms, but not for the other microturbulent velocities. These files have three columns: wavelength in $\AA$, a string of line identifications, and the equivalent width ($W$) in $\AA$. Weak lines with $W < 0.1 \ \AA$ are not listed. The new BOSZ grid contains 5\,186\,115 files, one for each of the eight resolutions for all 628\,620 spectra and one line id file per model atmosphere for the \vmicro \ = 0 \kms \ case, totaling 12.7~TB in size in gzip format.

\subsection{Differences between ATLAS9 and MARCS spectra}

\begin{figure}                          
\centering
\includegraphics[width=3.54in,angle=0]{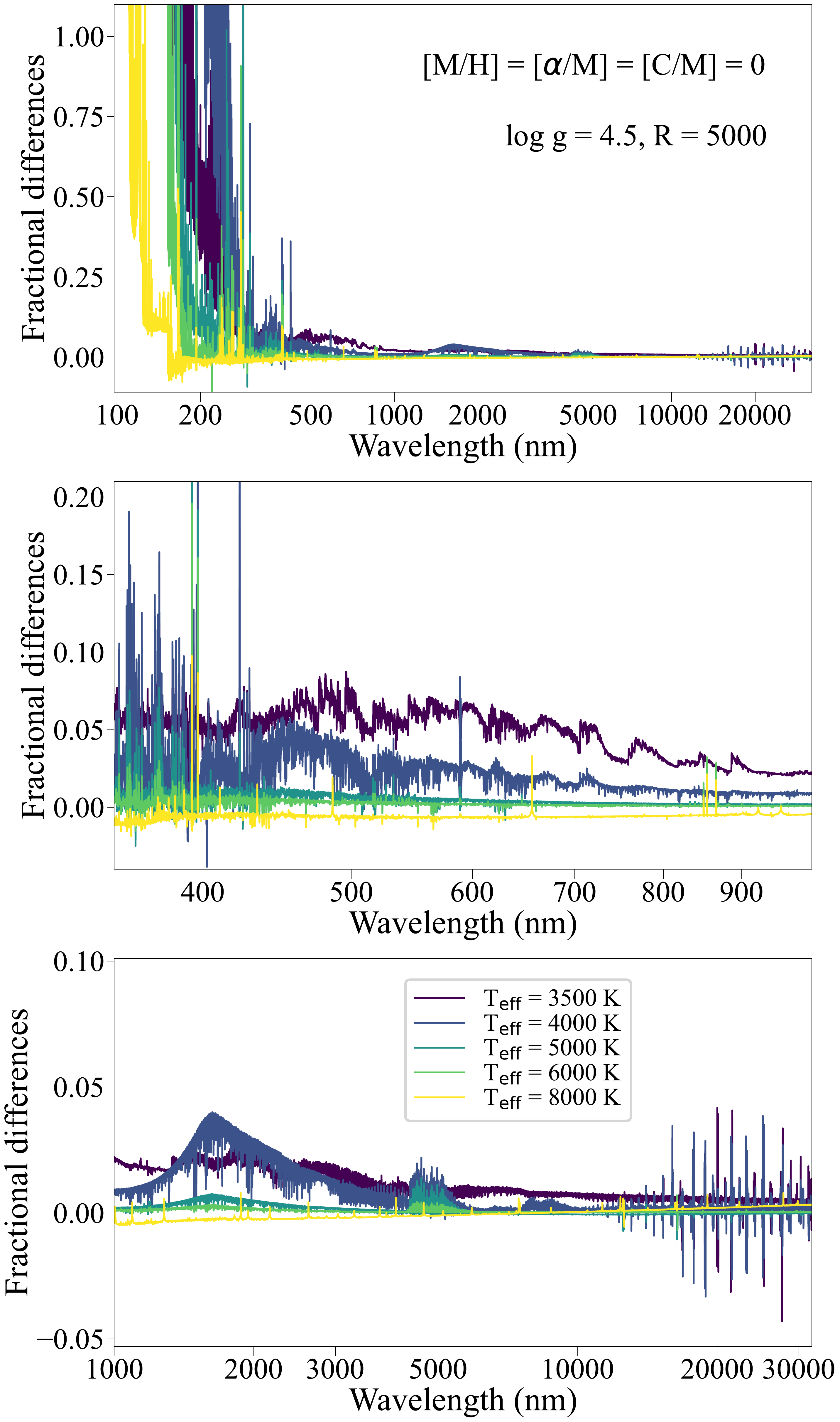}
\caption{Fractional differences between MARCS and ATLAS9 flux relative to the ATLAS9 flux. Top panel: Spectra shown from \teff \ = 3500~K to 8000~K and between 100 nm and 32 \mum. A large discrepancy exists in the UV for all temperatures. Middle panel: Relative differences in the optical and near-IR, where the offset is usually within 10\%. Bottom panel: Relative differences are the smallest in the IR between 1 and 32 \mum. The large spikes above 15 \mum \ originate from slight spectral mismatches in the resampling.}
\label{spectr}
\end{figure}

Between 7500 and 8000~K, both ATLAS9 and MARCS model atmospheres are published to quantify the difference between the two types of model atmospheres. MARCS replaces ATLAS9 models below 7500~K, but the differences due to the changes in model atmosphere structure are discussed before examining the changes between the new and old BOSZ grid. For this reason, four new spectra with ATLAS9 models at 3500, 4000, 5000, and 6000~K are calculated using the solar composition, \logg \ = 4.5, \vmicro \ = 2 \kms, and all the updates to the opacities. This set of four new cool ATLAS9 models quantifies the effects of changes in the atmospheric structures at lower temperatures. The ratio of differences between the MARCS and ATLAS9 spectra and the ATLAS9 spectra at $R$ = 5000 appear in the three panels of Figure~\ref{spectr}. The top panel shows the wavelength region between 100 nm and 32 \mum, the middle panel depicts the optical and near-IR between 350 and 1000~nm, and the bottom panel illustrates the IR region between 1000~nm and 32 \mum.

The difference of absolute fluxes between MARCS and ATLAS9 at 8000~K is on average $\pm$0.5\% at solar composition between 350 and 1000 nm and on average $\pm$0.3 \% between 1000~nm and 32 \mum . JWST covers 800~nm to 30~\mum . Larger than average differences in the optical and IR appear in the cores of very deep absorption lines where the flux itself is near zero. These ratios of small values appear as spikes in the panels of Figure \ref{spectr}, and are likely associated with a more limited extent of the MARCS models in the outermost atmospheric layers. The MARCS fluxes are also larger by 5$-$10\% than ATLAS9 for the coolest models of 3500 and 4000~K in the optical and near IR up to 5 \mum, but the differences decrease as the wavelength increases (middle and bottom panels of Figure~\ref{spectr}). Such systematic flux offsets in the optical and IR start to disappear at temperatures higher than 4000~K, and the agreement is within 2\% between 5 and 32 \mum. 

Differences increase substantially, at least to 10\% in the UV, for all temperatures between 3500 and 8000~K. In this wavelength region, the MARCS models have larger flux than the ATLAS9 models, and this discrepancy between the two types of models increases with decreasing wavelength. The MARCS UV excess reaches 8-10 times the ATLAS9 spectrum at the shortest wavelengths; but at such short wavelengths, the flux in both models is very low. This discrepancy is most likely the result of the continuous opacity changes in the UV region and the different model atmosphere structures.

\begin{figure*}                          
\centering
\includegraphics[width=7.1in,angle=0]{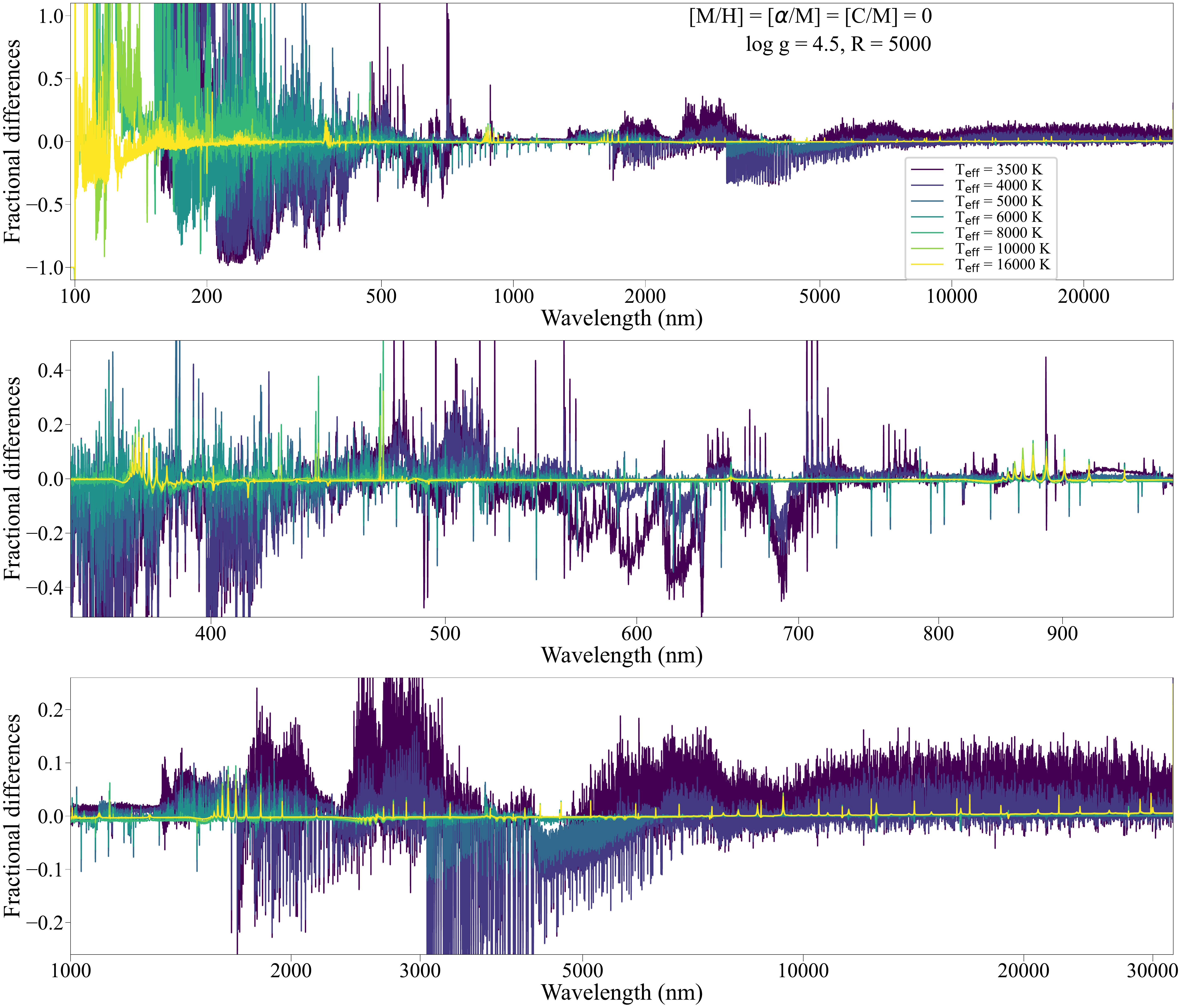}
\caption{Fractional differences between the flux of the new and old BOSZ grid relative to the old grid. Top panel: showing the full spectral range between 100 nm and 32 \mum. Middle panel: Relative differences in the optical range. Bottom panel: Relative differences in the IR.}
\label{spectr2}
\end{figure*}

\subsection{Description of file names}

File names are consistent, so the parameters of any spectra can be easily recovered. All new names starts with ``bosz2024\_'' and then the standard naming scheme follows. An example of naming a spectrum file with \teff \ = \ 5000~K, \logg \ = 5, \mh \ = 0, \vmicro \ = 0, $R$ = 500,  and solar scaled \am, \cm \ abundances is
$$\rm{mp\_t5000\_g+5.0\_m+0.00\_a+0.00\_c+0.00\_v0\_r500\_resam,}$$ where the first two strings define the model atmosphere, ``mp'': MARCS plan-parallel model, ``ms'': MARCS spherical model, ``ap'': ATLAS9 plan-parallel model. The ``\_t5000'' string indicates the effective temperature, which ranges between 2800 and 16000. The ``\_g+5.0'' string is the surface gravity that ranges between $-$0.5 and 5.5. The ``\_m-0.00'' denotes the metallicity covering $-$2.50 to 0.75. The ``\_a+0.00'' string shows the $\alpha$ abundance ($-$0.25 to 0.50), and the ``\_c+0.00'' string is the carbon abundance ($-$0.75 to 0.5). All the allowed values are listed in Tables~1 and 2.

After the strings indicating the main atmospheric parameters, the microturbulent velocity follows as ``\_v0,'' for which the allowed values are 0, 1, 2, and 4. The resolution is ``\_r500'' and allowed values are: 500, 1000, 2000, 5000, 10000, 20000, and 50000. For the original resolution, ``rorig'' appears here. The lower-resolution spectra were resampled, as indicated by the ``\_resam'' string. For the original resolution, the original wavelength sampling is kept, so this string is ``\_noresam.'' In the files containing the line identifications this part is called ``\_lineid.''

\section{Comparisons of the new and old BOSZ libraries}

\subsection{Changes in the model fluxes}

There are substantial changes in the fluxes of the new spectra when compared to the old BOSZ grid, as illustrated in Figure~\ref{spectr2}. The top panel shows the differences in flux relative to the old models in the common wavelength range of 100 nm and 32 \mum, the middle panel shows the optical, the bottom panel depicts the IR region. All comparisons are for temperatures in the common \teff \ range between 3500 and 16000~K using solar composition, \logg \ = 4.5, \vmicro \ = 2 \kms, and R = 5000. Two factors cause these changes: 1. the replacement of ATLAS9 with MARCS models, and 2. updates to the continuous and line opacities. The largest differences between the two sets of models appear in the UV. As seen in Section~3.4, differences in the UV between MARCS and ATLAS9 are substantial and probably on par with changes in opacities.

Except for the UV region, the differences in the model atmosphere structure modestly affects the overall flux only for the coolest stars below 4000~K. The large variations of flux displayed in Figure~\ref{spectr2} originate from the adoption of new continuous and line opacities. Changes in the continuous opacities mostly show up in the spectra of the \teff \ = 10000 and 16000~K stars (greenish yellow and yellow lines in the panels of Figure~\ref{spectr2}). Up to 10$-$15\% changes in the fluxes occur at these temperatures near the Balmer jump between 360 and 400~nm, and at 850 and 900~nm near the Paschen jump. The discrepancy in the hydrogen lines remains in the IR but decreases to about 5$-$7\%, at the most. 

\begin{scriptsize}
\begin{table*}
\caption{Atmospheric parameters of selected CALSPEC stars.}
\begin{tabular}{p{2.5cm}p{1.5cm}p{1.5cm}p{1.5cm}p{1.5cm}p{1.5cm}p{1.5cm}p{1.5cm}p{1.5cm}}
\hline
 &  & Old Grid &  &  &  & New Grid  &  &  \\
Star & \teff & \logg & \mh & \ebv & \teff & \logg & \mh & \ebv \\
\hline
BD+60 1753 & 9480  & 3.75 &  0.04 & 0.017  & 9420 & 3.75 &  0.12 & 0.015 \\
P330E      & 5830  & 4.80 & -0.21 & 0.028  & 5840 & 4.40 & -0.09 & 0.029 \\
KF06T2     & 4510  & 1.55 & -0.26 & 0.053  & 4550 & 2.10 & -0.11 & 0.069 \\
\end{tabular}
\label{modpar}
\end{table*}
\end{scriptsize}

\begin{figure}                          
\centering
\includegraphics[width=3.7in,angle=0,trim=70 70 0 30]{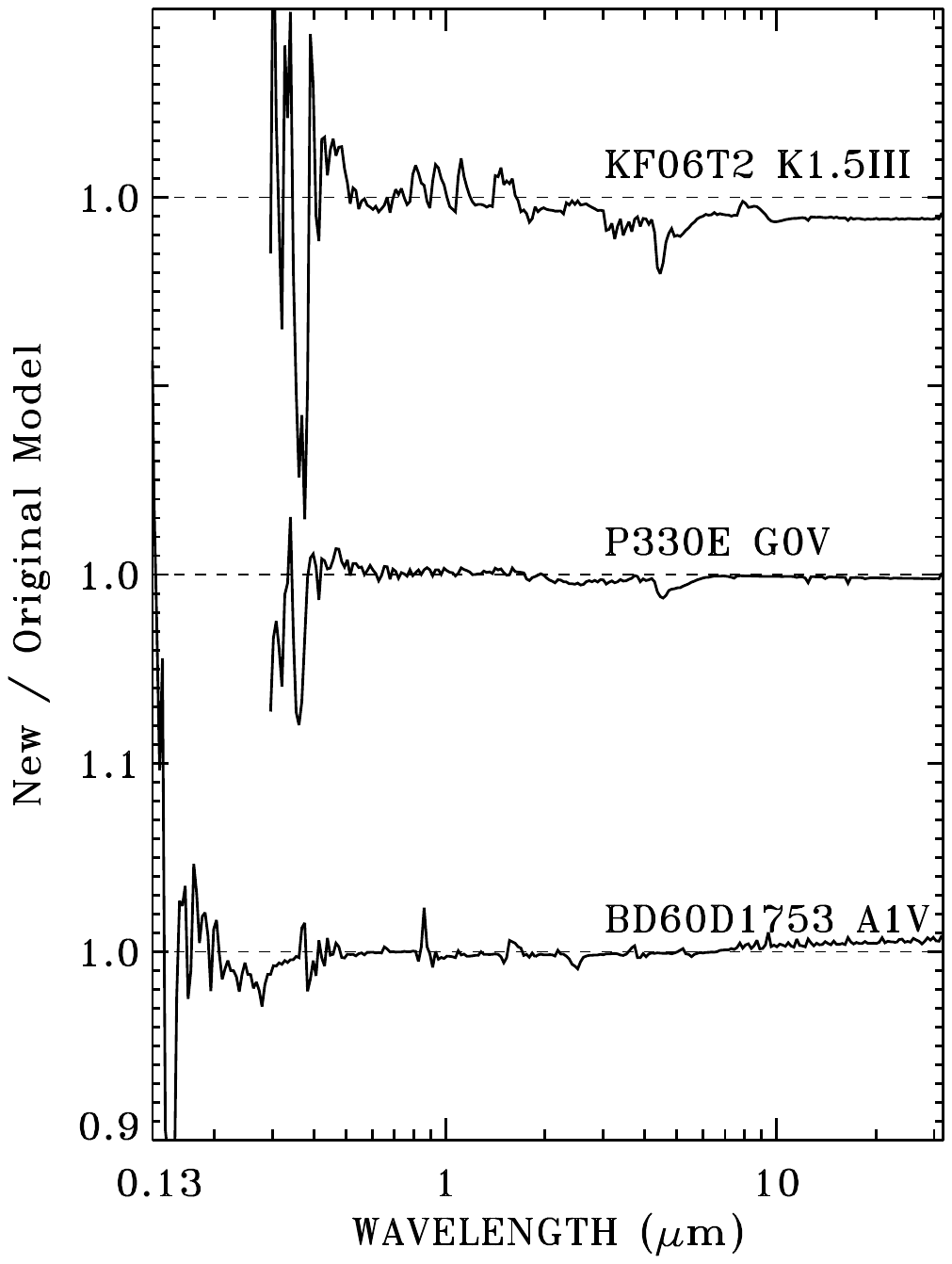}
\caption{\baselineskip=12pt
Ratio of the new to original best fitting models for an A-, G-, and K-type star. At the longer wavelengths, the old and new models agree mostly within 1\%. At shorter wavelengths, where the line blanketing is heavy, differences are much larger.}
\label{mzfig1}
\end{figure}

\begin{figure}                         
\centering
\includegraphics[width=3.7in,angle=0,trim=70 70 0 30]{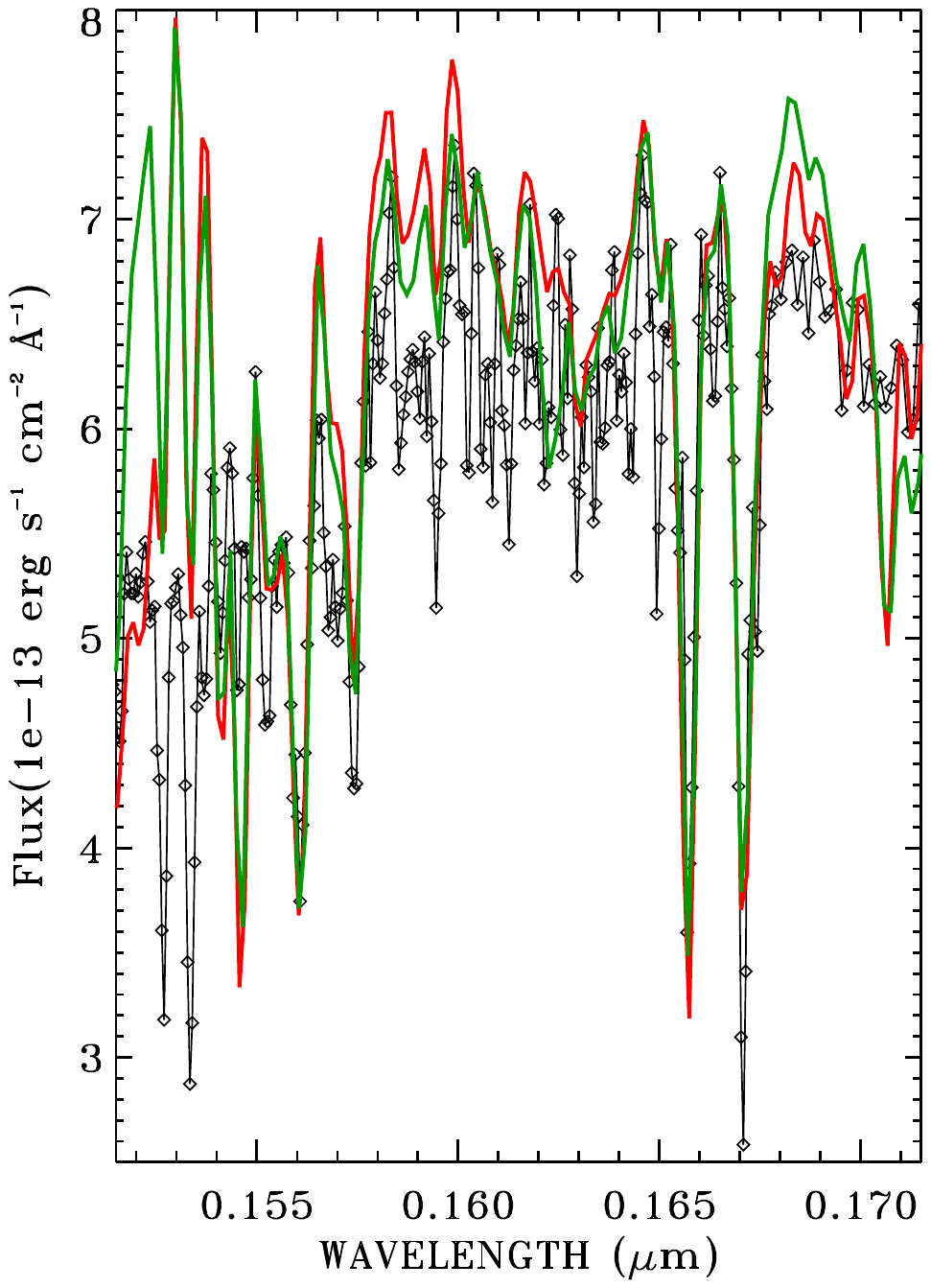}
\caption{SEDs for the STIS data (black diamonds), original model grid (green), and new model grid (red) in a region of heavy line blanketing for BD+60 1753.}
\label{mzfig2}
\end{figure}

For the coolest temperatures from 3500 to 5000~K (dark and light blue lines in Figure~\ref{spectr2}), the new BOSZ flux is significantly lower between 200 and 500 nm than the old grid. Some of this must be related to the addition of OH+, which has strong absorption bands in this wavelength region when \teff \ $<$ 5500~K (see Figure~\ref{molecules}) and which was not included in the old BOSZ models. Between 450 and 720~nm variations in flux from 20\% to $-$40\% appear at 3500~K and 4000~K, but not at hotter temperatures. The cause of this change is most likely the generally stronger absorption coming from the new TiO line list. Most of the large spikes of differences in the optical occur in the core of deep absorption lines, while the agreement is much better near the continuum.

The differences between the new and old BOSZ spectra are usually within $\pm$10\% when the wavelength is larger than 1~\mum. The discrepancy between the two versions of spectra increases as the temperature decreases, reaching a maximum at 3500~K. Because the largest variation only occurs in the spectra of the coolest models, the cause can be attributed to changes in the absorption bands of H$_2$O and the new addition of OH+. H$_2$O is a significant absorber above 1~\mum, and OH+ has absorption bands between 1.5 and 8~\mum. While the offset in fluxes is quite large when these molecules form in the atmosphere at \teff \ $<$ 5500~K, the agreement between the new and old models is generally better than 1\% when \teff \ $>$ 5500~K. Thus, models for CALSPEC stars cooler than 5500~K should have significant IR flux changes that increase with decreasing temperature when fitting with the new BOSZ models.

\subsection{Fitting CALSPEC stars}

To understand the typical differences between models that best fit the HST STIS and NICMOS observations, examples of an A-, G-, and K-type star are fit with both the new and the original grid using the technique of \citet{2017AJ....153..234B, 2019AJ....158..211B, 2020AJ....160...21B}. The model fitting finds the smallest $\chi^2$ by varying four parameters: $T_\mathrm{eff}$, $\log g$, [M/H], and E(B-V), which are the model effective temperature, surface gravity, metal abundance relative to hydrogen, and the selective extinction. The average galactic extinction prescription is from \citet{2022AJ....163..267G}.

Table~\ref{modpar} compares the model parameter results from fitting with the original and new model grids. There is fair agreement between the two sets of results with a maximum difference in \teff \ of 60~K for BD+60 1753. Figure~\ref{mzfig1} shows the ratios at a resolution R = 500 for new/old model SEDs that are defined by the parameters of the best fitting models of Table~\ref{modpar}. While model parameters may differ, differences in our final SEDs are less than 1\% over most of the longer wavelengths. Below the 0.29~\mum \ cutoff of the ratio for P330E and KF06T2, those stars are very faint.

Figure~\ref{mzfig2} depicts an example of the difficulties in modeling the data in a region of heavy line blanketing, where some features of the observed STIS SED (black diamonds) are not reproduced by either the new (red) or the original (green) model grid. To approximately match the STIS resolution, the models have R = 500. The strong doublet at 0.15267~\mum \ and 0.15334~\mum \ could be mostly interstellar SiII, which is not included in the models. The new grid matches the observation better at the shortest and longest wavelengths, while the original grid fits better in some intermediate  regions. Both grids are $>$7e-13 in comparison to the $\approx$5e-13 STIS flux at  0.1537~\mum. Thus, for applications in the observed STIS wavelength range, the observed CALSPEC flux is preferred to any model.

\section{Conclusions}

The new set of BOSZ models, available from MAST\footnote{https://archive.stsci.edu/prepds/bosz} at the Space Telescope Science Institute, contains 628\,620 model spectra calculated with Synspec using an LTE approximation. The spectra cover a wavelength range between 50 nm and 32 \mum, using eight different resolutions from 500 to roughly 200\,000, depending on the \teff. Spectra with varying metallicities from $-$2.5 to 0.75~dex, \am \ from $-$0.25 to 0.5~dex, and \cm \ from $-$0.75 to 0.5~dex have been calculated. The hottest spectra in the new grid are at \teff \ = 16000~K using ATLAS9 model atmospheres; the inclusion of MARCS models below 8000~K extends the grid down to 2800~K, where 23 molecular line lists, many from the ExoMol project, play an important role in shaping the IR spectrum. 

As a result of using MARCS models at lower temperatures and updates in the continuous and line opacities, there are large differences between the new and old BOSZ grid in the common wavelengths of 100 nm and 32 \mum. Above 10000~K, flux changes of up to 10$-$15\% occur near the Balmer jump between 360 and 400~nm and at 850 and 900~nm near the Paschen jump, while the IR hydrogen lines have smaller differences than in the optical. In the IR, variations occur in the spectra of the coolest models, with changes of up to $\pm$10\%, which are mostly due to absorption bands of a new H$_2$O line list from ExoMol and to the new addition of OH+. Due to the updated molecular line lists, CALSPEC stars cooler than 5500~K have significant IR flux changes compared to the old BOSZ spectra, which increase with decreasing temperature.

Using both the new and old BOSZ grid, the best fitting models for an A-, G-, or K-type star from the CALSPEC database illustrate the typical differences between the three pairs of resulting SEDs. The main parameters, \teff, \logg, and \mh , agree well; as expected from the direct model comparisons, the largest differences can be seen in the UV region. While none of the models can exactly reproduce all observed UV features, the new models match the observations slightly better than the old ones.

\begin{acknowledgements}
We thank the referee for his/her comments that significantly improved the paper. This project has been supported by the LP2021-9 Lend\"ulet grant of the Hungarian Academy of Sciences. On behalf of the "Calculating the Synthetic Stellar Spectrum Database of the \textit{James Webb} Space Telescope" project we are grateful for the possibility to use HUN-REN Cloud (see \citealt{heder2022}; https://science-cloud.hu/) which helped us achieve the results published in this paper. B.Cs. acknowledges support from the Lend\"ulet Program LP2023-10 of the Hungarian Academy of Sciences. C.A.P. acknowledges financial support from the Spanish Ministry MICINN projects AYA2017-86389-P and PID2020-117493GB-I00.

Funding for the Sloan Digital Sky Survey IV has been provided by the Alfred P. Sloan Foundation, the U.S. Department of Energy Office of 
Science, and the Participating Institutions. 

SDSS-IV acknowledges support and resources from the Center for High Performance Computing  at the University of Utah. The SDSS website is www.sdss.org.

SDSS-IV is managed by the Astrophysical Research Consortium for the Participating Institutions of the SDSS Collaboration including the Brazilian Participation Group, the Carnegie Institution for Science, Carnegie Mellon University, Center for Astrophysics | Harvard \& Smithsonian, the Chilean Participation Group, the French Participation Group, Instituto de Astrof\'isica de Canarias, The Johns Hopkins University, Kavli Institute for the Physics and Mathematics of the Universe (IPMU) / University of Tokyo, the Korean Participation Group, Lawrence Berkeley National Laboratory, Leibniz Institut f\"ur Astrophysik Potsdam (AIP),  Max-Planck-Institut f\"ur Astronomie (MPIA Heidelberg), Max-Planck-Institut f\"ur Astrophysik (MPA Garching), Max-Planck-Institut f\"ur Extraterrestrische Physik (MPE), National Astronomical Observatories of China, New Mexico State University, New York University, University of Notre Dame, Observat\'ario Nacional / MCTI, The Ohio State University, Pennsylvania State University, Shanghai Astronomical Observatory, United Kingdom Participation Group, Universidad Nacional Aut\'onoma de M\'exico, University of Arizona, University of Colorado Boulder, University of Oxford, University of Portsmouth, University of Utah, University of Virginia, University of Washington, University of Wisconsin, Vanderbilt University, and Yale University.

\end{acknowledgements}

\bibliographystyle{aa}
\bibliography{references} 
\end{document}